\documentclass[aps,nofootinbib,superscriptaddress, showpacs,preprintnumbers,twocolumn]{revtex4-2}

\usepackage{epsfig}
\usepackage{multirow}
\usepackage{subcaption}
\usepackage{eurosym}
\usepackage{dcolumn}
\usepackage{bm}
\usepackage{enumerate}
\usepackage{float}
\usepackage{epstopdf}
\usepackage{amsmath}
\usepackage{amsthm}
\usepackage{xcolor}
\usepackage{amsthm}
\newtheorem{theorem}{Theorem}
\newtheorem{corollary}{Corollary}
\newtheorem{lemma}{Lemma}
\usepackage{bm}
\usepackage{amsfonts}
\usepackage{amssymb}
\usepackage{graphicx}
\usepackage{alphalph,mathtools}
\usepackage{etoolbox}
\usepackage{amsthm}
\usepackage{color}
\usepackage{booktabs}
\usepackage{hyperref}
\hypersetup{colorlinks,citecolor=blue}
\usepackage{footnote}
\usepackage{makecell,tabularx}

\def\be{\begin{equation}}
\def\ee{\end{equation}}
\def\bea{\begin{eqnarray}}
\def\eea{\end{eqnarray}}

\begin{document}
\title{ Gravitational Light Deflection via Fractional Calculus and Chebyshev Polynomial Approximation } 

\author{Aniruddha Ghosh}
 \email{ruddha.g@gmail.com}
 \affiliation{%
 Department of Mathematics, Indian Institute of Engineering Science and Technology, Shibpur, Howrah-711103, India.\\ 
}%
\author{Ujjal Debnath}%
 \email{ujjaldebnath@gmail.com}
\affiliation{%
 Department of Mathematics, Indian Institute of Engineering Science and Technology, Shibpur, Howrah-711103, India.\\ 
}%

\begin{abstract}
We develop a fractional-calculus-based framework for evaluating gravitational light deflection in static, spherically symmetric spacetimes. By introducing a suitable transformation of the radial coordinate, the deflection integral is recast into an Abel-type integral equation, from which the associated kernel is reconstructed using fractional integration. The leading singular behaviour near the photon sphere is isolated analytically, and the remaining regular part is approximated by a finite Chebyshev polynomial expansion. This combination retains the correct strong-field behaviour while providing an efficient numerical representation of the deflection integral. We demonstrate the accuracy of the method using the Schwarzschild spacetime as a benchmark and compare the resulting deflection angles with the standard strong-deflection approximation and a two-term weak-deflection expansion. We further apply the method to calculate strong-field lensing observables, including the limiting angular position, angular separation, and relative magnification of relativistic images, for the supermassive black holes Sgr A*. The results show that the proposed approach provides an accurate and flexible means of evaluating gravitational deflection angles across a broad range of closest-approach radii. The framework can be extended to other static, spherically symmetric black-hole geometries, including solutions arising in modified theories of gravity.
\par
\vspace{0.1cm}
\end{abstract}
\maketitle
\section{Introduction}\label{sec1}
Gravitational lensing is one of the important predictions of general relativity and provides a powerful means of investigating the geometry of spacetime in the vicinity of compact objects. According to general relativity, the gravitational field of a massive body curves spacetime, causing the propagation of light to deviate from its otherwise straight-line trajectory. Although the deflection of light in the weak-field regime has been extensively investigated, the strong-field regime offers a richer structure in which photons can undergo substantial deflection before reaching a distant observer. In particular, when a light ray passes sufficiently close to a black hole, its trajectory can approach an unstable circular null orbit, commonly known as the photon sphere. As a result, the deflection angle increases without bound as the closest-approach radius approaches the photon sphere, giving rise to an infinite sequence of relativistic images \cite{Einstein1936,Darwin1959,VirbhadraEllis2000}.

The study of strong-field gravitational lensing is important for understanding the spacetime geometry around black holes and for extracting information about their physical properties from observable lensing signatures. Several analytical approaches have been developed to describe the propagation of light rays in the strong-field regime. Among them, the strong-deflection-limit formalism provides a particularly useful description of the logarithmic divergence of the deflection angle near the photon sphere \cite{Bozza2001,Bozza2002}. This approach has been widely applied to different black-hole spacetimes and has enabled the calculation of observables associated with relativistic images \cite{VirbhadraEllis2000,Perlick2004,IyerPetters2007}. However, the standard strong-deflection expansion is an asymptotic approximation constructed for trajectories close to the photon sphere. Its accuracy may therefore deteriorate when it is applied to trajectories with larger closest-approach radii. This motivates the development of alternative methods capable of evaluating the deflection angle accurately over a broader range of impact parameters.

In this work, we develop a numerical framework for gravitational light deflection by combining fractional calculus with Chebyshev polynomial approximation. We consider a general static, spherically symmetric spacetime described by
\begin{equation}
ds^2=-A(r)dt^2+B(r)dr^2+C(r)d\Omega^2,
\end{equation}
where A(r), B(r), and C(r) are functions of the radial coordinate. Starting from the null geodesic equations, the deflection angle is expressed as an integral involving the metric functions and the closest-approach radius. By introducing a suitable transformation of the radial coordinate, we recast this integral into an Abel-type integral equation. The corresponding kernel can then be reconstructed using fractional integration \cite{GorenfloVessella1991,Podlubny1999}.

A central difficulty in evaluating the resulting integral is the singular behaviour of its kernel near the photon sphere. Rather than approximating the complete kernel directly, we isolate its leading singular contribution analytically and subtract it to obtain a regularised function. The regularised part is subsequently approximated by a finite expansion in Chebyshev polynomials. The Chebyshev--Gauss nodes provide an efficient means of determining the expansion coefficients, while the analytically extracted singular term preserves the correct behaviour in the strong-field limit. This procedure combines the advantages of an analytical treatment near the photon sphere with an accurate numerical approximation over a finite radial interval. Chebyshev polynomial expansions are particularly useful in this context because of their favourable approximation properties and their widespread application in spectral numerical methods \cite{Boyd2001,Trefethen2013}.

To examine the accuracy and applicability of the proposed framework, we consider the Schwarzschild black hole as a benchmark. We compare the resulting deflection angles with those obtained from the standard strong-deflection-limit approximation and a two-term weak-deflection expansion. We then use the calculated deflection angle to investigate strong-field lensing observables, including the limiting angular position of relativistic images, their angular separation, and their relative magnifications. As illustrative applications, we consider the supermassive black holes Sgr A* \cite{melia2001supermassive}.

The remainder of this paper is organised as follows. In Sec.\eqref{sec2}, we present the general formulation of gravitational light deflection in static, spherically symmetric spacetimes. In Sec. \eqref{Sec3}, we develop the fractional-calculus and Chebyshev polynomial framework for evaluating the deflection integral, including the analytical treatment of the singular behaviour near the photon sphere. In Sec.\eqref{Sec4}, we apply the proposed method to the Schwarzschild black hole and examine its numerical accuracy by comparing the resulting deflection angles with standard strong- and weak-field approximations. In Sec.\eqref{sec5}, we investigate the strong-field lensing observables derived from the Chebyshev deflection angle, with particular applications to Sgr A*. Finally, Sec.\eqref{Sec6}summarises our main results and discusses possible directions for future research.

\section{Construction}\label{sec2}
We first briefly discuss the construction of the light-deflection angle for the general static, spherically symmetric spacetime.
We first consider a general static, spherically symmetric spacetime described by the line element
\begin{equation}\label{e1}
ds^{2}=-A(r)\,dt^{2}+B(r)\,dr^{2}+C(r)\,d\Omega^{2},
\end{equation}
where
\begin{equation}\label{e2}
d\Omega^{2}=d\theta^{2}+\sin^{2}\theta\,d\phi^{2},
\end{equation}
and $A(r)$, $B(r)$, and $C(r)$ are functions of the radial coordinate $r$.
Owing to spherical symmetry, the photon trajectory can be restricted, without loss of generality, to the equatorial plane, $\theta=\pi/2$. Consequently, $d\theta=0$ and $\sin^{2}\theta=1$, such that
\begin{equation}\label{e4}
d\Omega^{2}=d\phi^{2}.
\end{equation}
The motion of a photon in a curved spacetime is governed by the null geodesic equations. A convenient way to obtain these equations is through the variational formulation of geodesic motion. For a spacetime characterized by the metric tensor $g_{\mu\nu}$, the geodesic Lagrangian is defined as
\begin{equation}\label{e5}
\mathcal{L}
=\frac{1}{2}g_{\mu\nu}\dot{x}^{\mu}\dot{x}^{\nu},
\end{equation}
where $x^{\mu}$ denotes the spacetime coordinates and a dot represents differentiation with respect to an affine parameter $\lambda$, i.e.,
$\dot{x}^{\mu}\equiv\frac{dx^{\mu}}{d\lambda}.$
For the static, spherically symmetric spacetime considered above, and restricting the motion to the equatorial plane $\theta=\pi/2$, Eq.~\eqref{e4} takes the explicit form
\begin{equation}\label{e6}
2\mathcal{L}
=-A(r)\dot{t}^{\,2}
+B(r)\dot{r}^{\,2}
+C(r)\dot{\phi}^{\,2}.
\end{equation}
Since the trajectory under consideration corresponds to a photon, the spacetime interval along the trajectory vanishes, $ds^{2}=0$. Consequently, the null condition requires
\begin{equation}
2\mathcal{L}
=-A(r)\dot{t}^{\,2}
+B(r)\dot{r}^{\,2}
+C(r)\dot{\phi}^{\,2}=0.
\label{e7}
\end{equation}
The physical trajectory is obtained by extremizing the corresponding action,
\begin{equation}
S=\int \mathcal{L}\,d\lambda.
\label{e8}
\end{equation}
Applying the principle of stationary action, $\delta S=0$, leads to the Euler--Lagrange equations for each generalized coordinate $q^{\mu}\in\{t,r,\phi\}$,
\begin{equation}
\frac{d}{d\lambda}
\left(
\frac{\partial\mathcal{L}}
{\partial\dot{q}^{\mu}}
\right)
-
\frac{\partial\mathcal{L}}
{\partial q^{\mu}}
=0.
\label{e9}
\end{equation}
These equations are equivalent to the geodesic equations and completely determine the photon trajectory once the appropriate initial conditions are specified \cite{Chandrasekhar1983,Perlick2004}. Moreover, because the metric coefficients depend only on the radial coordinate $r$, the coordinates $t$ and $\phi$ are cyclic coordinates. Their corresponding Euler--Lagrange equations therefore give rise to two conserved quantities along the photon trajectory, which may be identified with the conserved energy and angular momentum, respectively. These constants of motion will be used below to obtain the radial equation and, subsequently, the expression for the light-deflection angle.

Since the metric functions $A(r)$, $B(r)$, and $C(r)$ depend only on the radial coordinate $r$, the Lagrangian has no explicit dependence on $t$ and $\phi$. Hence, $t$ and $\phi$ are cyclic coordinates. According to Noether's theorem, the corresponding time-translation and rotational symmetries lead to the conservation of the photon energy $E$ and angular momentum $L$, respectively. The associated conserved quantities are
\begin{equation}
E=A(r)\dot{t},
\qquad
L=C(r)\dot{\phi},
\label{e10}
\end{equation}
or, equivalently,
\begin{equation}
\dot{t}=\frac{E}{A(r)},
\qquad
\dot{\phi}=\frac{L}{C(r)}.
\label{e11}
\end{equation}
Substituting Eq.~\eqref{e10} into the null geodesic condition given by Eq.~\eqref{e6}, 
we obtain the radial equation of motion for the photon as
\begin{equation}
B(r)\dot{r}^{\,2}
=
\frac{E^{2}}{A(r)}
-
\frac{L^{2}}{C(r)}.
\label{e12}
\end{equation}
Using $\dot{r}$ and $\dot{\phi}$, the variation of the azimuthal angle 
$\phi$ with respect to the radial coordinate $r$ is given by
\begin{equation}\label{e13}
\frac{d\phi}{dr}
=
\frac{\dot{\phi}}{\dot{r}}
=
\frac{L\sqrt{B(r)}}{C(r)}
\left[
\frac{E^{2}}{A(r)}
-
\frac{L^{2}}{C(r)}
\right]^{-1/2}.
\end{equation}
Introducing the impact parameter
\begin{equation}\label{e14}
b\equiv\frac{L}{E},
\end{equation}
and factoring $E^{2}$ from the term inside the square root, the above
expression reduces to
\begin{equation}
\frac{d\phi}{dr}
=
\frac{b\sqrt{B(r)}}{C(r)}
\left[
\frac{1}{A(r)}
-
\frac{b^{2}}{C(r)}
\right]^{-1/2}.
\label{e15}
\end{equation}
At the point of closest approach, $r=r_{0}$, the radial velocity of the 
photon vanishes, i.e., $\dot{r}=0$. Applying this condition to the radial 
equation of motion, we obtain the relation between the impact parameter 
$b$ and the closest-approach distance $r_{0}$ as
\begin{equation}
b^{2}
=
\frac{L^{2}}{E^{2}}
=
\frac{C(r_{0})}{A(r_{0})}.
\label{e16}
\end{equation}
As the photon approaches the compact object from infinity, the radial
coordinate decreases from $r=\infty$ to the point of closest approach
$r=r_{0}$, corresponding to the incoming branch of the trajectory.
After reaching $r_{0}$, the photon moves outward from $r=r_{0}$ to
$r=\infty$, which represents the outgoing branch. Owing to the symmetry
of the trajectory about the point of closest approach, the magnitudes
of the angular changes along the incoming and outgoing branches are
equal. Therefore, the total change in the azimuthal angle is
\begin{equation}
\Delta\phi
=
\Delta\phi_{\rm in}
+
\Delta\phi_{\rm out}
=
2\int_{r_{0}}^{\infty}
\frac{d\phi}{dr}dr
\equiv {\mathcal I(r_{0})}.
\label{e17}
\end{equation}
Therefore, the deflection angle of the photon is obtained by subtracting
the corresponding angular change $\pi$ for an undeflected trajectory
from the total change in the azimuthal angle. Thus,
\begin{equation}
\mathcal{A}(r_{0})=\mathcal{I}(r_{0})-\pi.
\label{e18}
\end{equation}
\begin{figure}
    \centering
    \includegraphics[width=1.2 \linewidth]{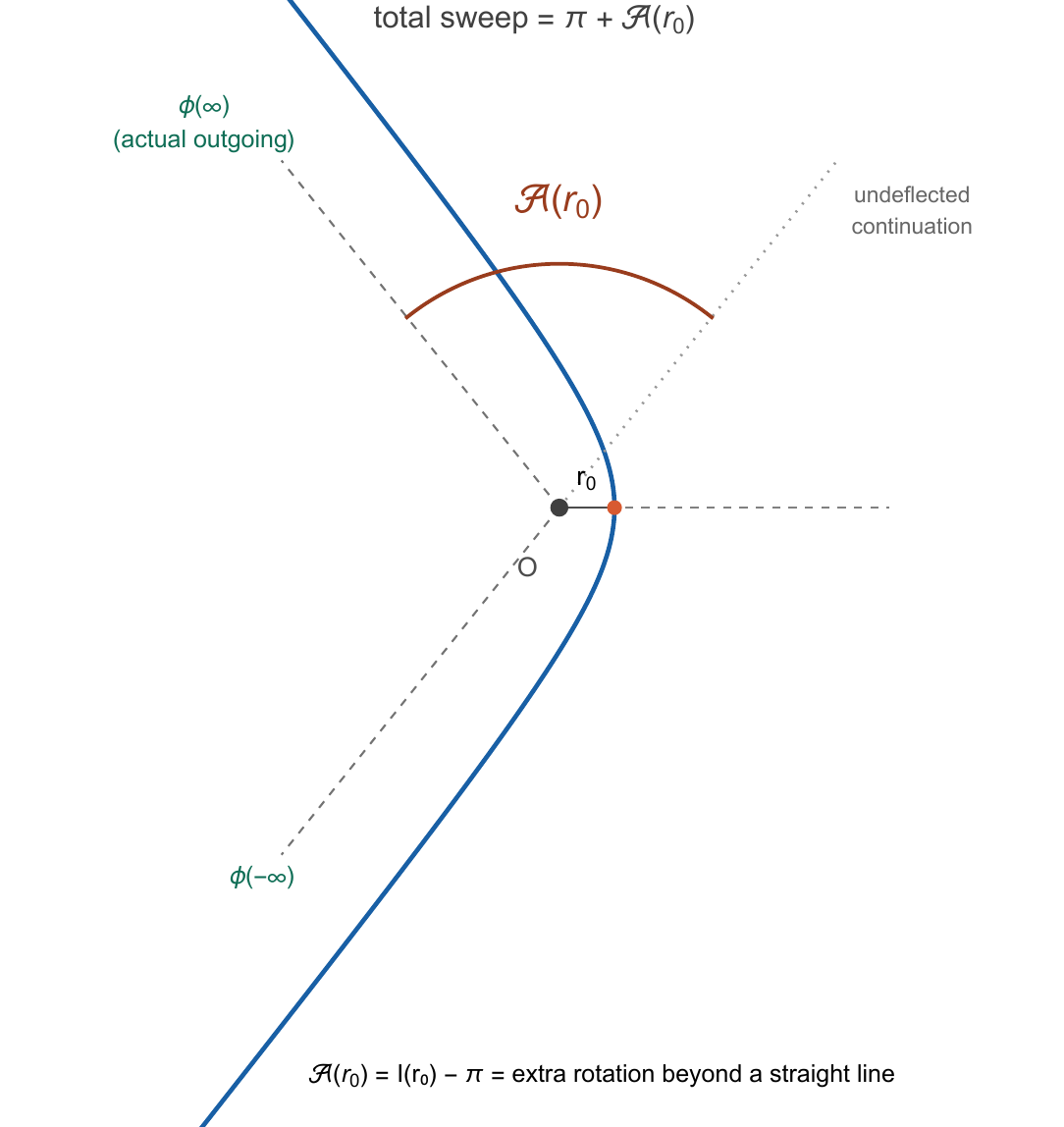}
    \caption{The photon approaches the compact object from infinity, reaches its closest-approach distance $r_0$, and subsequently escapes to infinity.}
    \label{f1}
\end{figure}
where $\mathcal{A}(r_{0})$ denotes the deflection angle, while the total
change in the azimuthal angle is given by
\begin{equation}
\mathcal{I}(r_{0})
=
2\int_{r_{0}}^{\infty}
\frac{\sqrt{B(r)}}{\sqrt{C(r)}}
\left[
\frac{C(r)A(r_{0})}{C(r_{0})A(r)}
-1
\right]^{-1/2}
dr.
\label{e19}
\end{equation}
In principle, once the integral $\mathcal{I}(r_{0})$ is evaluated, the
corresponding deflection angle $\mathcal{A}(r_{0})$ can be determined
directly \cite{Darwin1959,Bozza2002,Perlick2004}. However, an analytical
evaluation of this integral is, in general, nontrivial. In particular,
the integrand exhibits a singular behavior at the lower limit $r=r_{0}$, while the upper integration limit
extends to infinity. Various approximation schemes have been developed
to address these difficulties, including the weak-deflection expansion
and the formalism introduced by Bozza \cite{Bozza2002,IyerPetters2007}. Although these approaches provide
powerful tools for studying gravitational deflection, the accuracy of an
approximate evaluation of the integral and the associated truncation
error are not always controlled explicitly within the approximation
itself. Motivated by this issue, in the following section we develop an
alternative approximation scheme based on fractional calculus together
with a Chebyshev--minimax treatment. This construction is intended to
provide a systematic approximation of the deflection integral while
allowing the corresponding approximation error to be quantified and
controlled.
\section{Fractional--Chebyshev Treatment of the Deflection Integral}\label{Sec3}
In this section, we develop an approximation scheme for evaluating the
deflection angle given by Eq.~\eqref{e18}, employing fractional calculus together
with a Chebyshev--minimax treatment. Since the deflection angle is expressed
in terms of $\mathcal{I}(r_{0})$, the problem reduces to obtaining an
appropriate approximation for the integral $\mathcal{I}(r_{0})$.
\begin{theorem}
The integral $\mathcal{I}(r_{0})$ governing the deflection angle admits a
representation in terms of a Weyl fractional integral.
\end{theorem}
\begin{proof}
We begin by introducing the function
\begin{equation}
h(r)=\frac{C(r)}{A(r)}\equiv w.
\label{e20}
\end{equation}
At the point of closest approach, $r=r_{0}$, we define
\begin{equation}\label{e21}
h(r_{0})=w_{0}.
\end{equation}
Using the relation between the impact parameter and the closest-approach
distance, we immediately obtain
\begin{equation}
w_{0}=h(r_{0})
=\frac{C(r_{0})}{A(r_{0})}
=b^{2}.
\label{e22}
\end{equation}
We now employ the transformation $w=h(r)$ to recast the deflection
integral into a form suitable for a Weyl fractional-integral
representation.\cite{Perlick2004} We now change the integration variable from $r$ to $w$ by using
$w=h(r)$. Accordingly,
\begin{equation}\label{e23}
dw=h'(r)\,dr,
\qquad
dr=\frac{dw}{h'(r)}.
\end{equation}
The transformation is well defined in the region where $h'(r)\neq 0$,
i.e., away from the photon sphere. Under this transformation,
$r=r_{0}$ corresponds to $w=w_{0}$. The integral
$\mathcal{I}(r_{0})$ can therefore be recast as
\begin{equation}
\mathcal{I}(w_{0})
=
2\sqrt{w_{0}}
\int_{w_{0}}^{\infty}
\underbrace{
\frac{\sqrt{B(r(w))}}
{\sqrt{C(r(w))}\,h'(r(w))}
}_{\displaystyle G(w)}
(w-w_{0})^{-1/2}\,dw.
\label{e24}
\end{equation}
Introducing
\begin{equation}
G(w)
\equiv
\frac{\sqrt{B(r(w))}}
{\sqrt{C(r(w))}\,h'(r(w))},
\label{e25}
\end{equation}
the above expression takes the compact form
\begin{equation}
\mathcal{I}(w_{0})
=
2\sqrt{w_{0}}
\int_{w_{0}}^{\infty}
G(w)(w-w_{0})^{-1/2}\,dw.
\label{e26}
\end{equation}
To identify the fractional structure of Eq.~(\ref{e26}), we recall
the right-sided Weyl fractional integral.\cite{Samko1993,GorenfloVessella1991} Let $\alpha>0$ and let
$G:[x,\infty)\rightarrow\mathbb{R}$ be such that the following integral
exists. The right-sided Weyl fractional integral of order $\alpha$ is
defined by
\begin{equation}
\left(I_{-}^{\alpha}G\right)(x)
:=
\frac{1}{\Gamma(\alpha)}
\int_{x}^{\infty}
(t-x)^{\alpha-1}G(t)\,dt,
\qquad x>0,
\label{e27}
\end{equation}
where $\Gamma(\alpha)$ is the Euler Gamma function. In particular, for
$0<\alpha<1$, the kernel $(t-x)^{\alpha-1}$ possesses an integrable
singularity at the lower limit $t=x$, provided that $G$ is sufficiently
regular in a neighbourhood of $x$. Convergence at the upper limit further
requires an appropriate decay of $G(t)$ as $t\rightarrow\infty$.
Setting
\begin{equation}\label{e28}
\alpha=\frac{1}{2},\qquad x=w_{0},\qquad t=w,
\end{equation}
and using
\begin{equation}\label{e29}
\Gamma\left(\frac{1}{2}\right)=\sqrt{\pi},
\end{equation}
the Weyl fractional integral becomes
\begin{equation}
\left(I_{-}^{1/2}G\right)(w_{0})
=
\frac{1}{\sqrt{\pi}}
\int_{w_{0}}^{\infty}
(w-w_{0})^{-1/2}G(w)\,dw.
\label{e30}
\end{equation}
Comparison with Eq.~(\ref{e26}) therefore yields
\begin{equation}
\boxed{
\mathcal{I}(w_{0})
=
2\sqrt{\pi w_{0}}\,
\left(I_{-}^{1/2}G\right)(w_{0})
}.
\label{e31}
\end{equation}
Hence, the deflection integral $\mathcal{I}$ is represented by a
right-sided Weyl fractional integral of order $1/2$, completing the
proof.
\end{proof}
\begin{corollary}
The Weyl fractional representation of the deflection integral admits
an inversion formula for reconstructing the metric-dependent
function $G(w)$ from the deflection angle.
\end{corollary}

\begin{proof}
The right-sided Weyl fractional integrals satisfy the semigroup property
\cite{Samko1993,Podlubny1999}
\begin{equation}
I_{-}^{\mu}I_{-}^{\nu}
=
I_{-}^{\mu+\nu},
\label{e32}
\end{equation}
provided that the corresponding operators and integrals are well defined.
In particular,
\begin{equation}
I_{-}^{1/2}I_{-}^{1/2}=I_{-}^{1},
\label{e33}
\end{equation}
where $I_{-}^{1}$ reduces to ordinary right-sided integration.

From the fractional representation obtained above, we define
\begin{equation}
\Phi(w_{0})
:=
\frac{\mathcal{I}(w_{0})}
{2\sqrt{\pi w_{0}}},
\label{e34}
\end{equation}
so that
\begin{equation}
\Phi(w_{0})
=
\left(I_{-}^{1/2}G\right)(w_{0}).
\label{e35}
\end{equation}
Introducing the right-sided Weyl fractional derivative
$D_{-}^{1/2}$, which acts as a left inverse of
$I_{-}^{1/2}$ under the appropriate regularity and decay
conditions, we apply it to Eq.~(\ref{e35}) to obtain
\begin{equation}
G(w_{0})
=
\left(D_{-}^{1/2}\Phi\right)(w_{0}).
\label{e36}
\end{equation}
For sufficiently regular and decaying $\Phi$, the right-sided
Weyl fractional derivative of order $1/2$ is given by
\cite{Samko1993,Podlubny1999}
\begin{equation}
G(w_{0})
=
-\frac{1}{\sqrt{\pi}}
\frac{d}{dw_{0}}
\int_{w_{0}}^{\infty}
\frac{\Phi(t)}
{\sqrt{t-w_{0}}}\,dt.
\label{e37}
\end{equation}
Since
\begin{equation}
\mathcal{I}(w_{0})
=
\mathcal{A}(w_{0})+\pi,
\label{e38}
\end{equation}
the function $\Phi$ can be determined once the deflection angle
$\mathcal{A}$ is known. Equation~(\ref{e37}) therefore provides
an inversion of the deflection integral, yielding
\begin{equation}
G(w)
=
\frac{\sqrt{B(r(w))}}
{\sqrt{C(r(w))}\,h'(r(w))},
\qquad
h(r)=\frac{C(r)}{A(r)}.
\label{e39}
\end{equation}
Thus, knowledge of the deflection angle as a function of the
closest-approach parameter determines the corresponding
metric-dependent combination $G(w)$ through fractional inversion,
without requiring a weak- or strong-deflection expansion.
\end{proof}
The inverse problem formulated above was previously developed and solved
in our earlier work by Ghosh, Debnath, \textit{et al.}~\cite{ghosh2026abel}.

\begin{figure}
    \centering
    \includegraphics[width=1.2\linewidth]{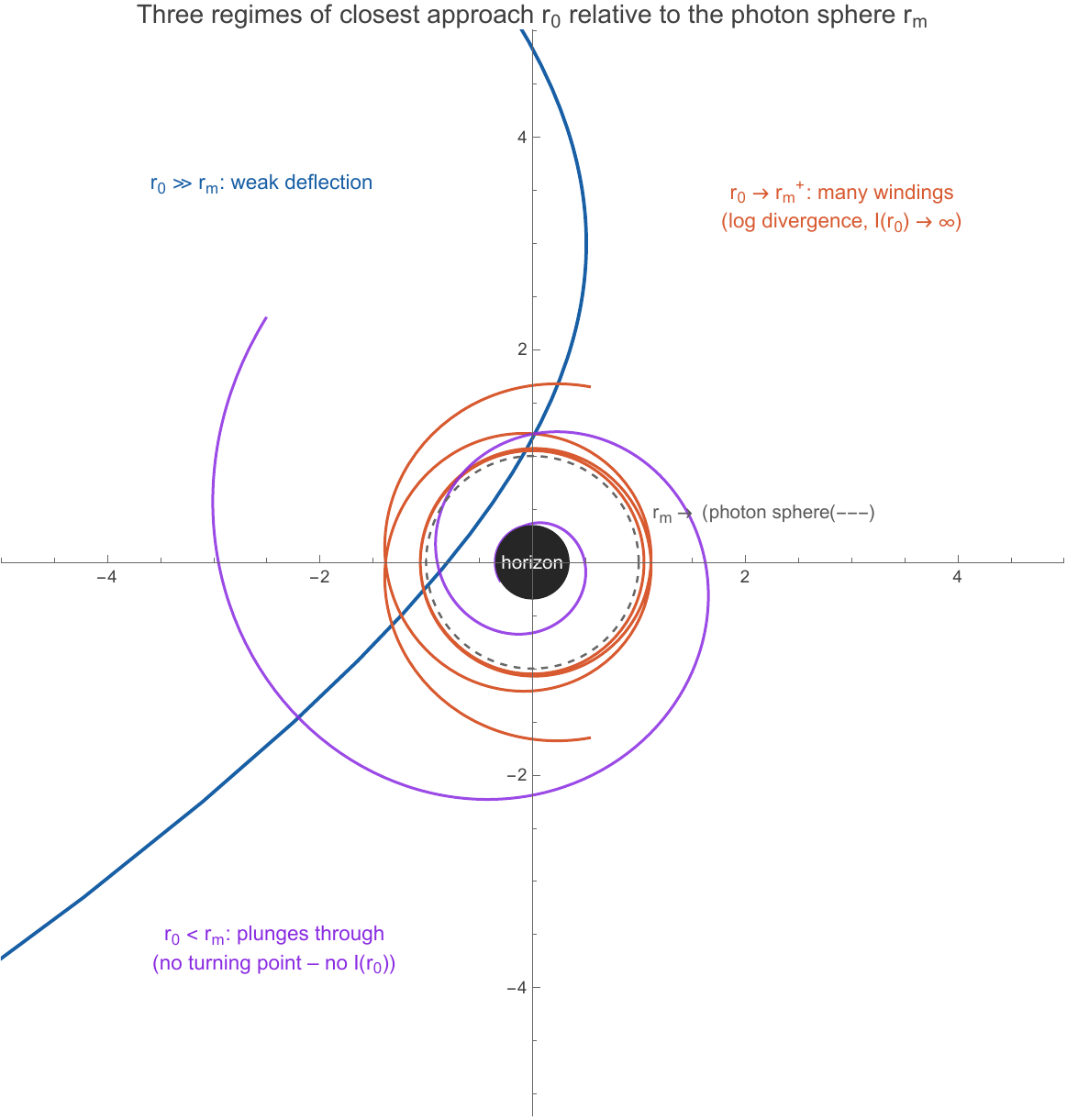}
    \caption{Schematic illustration of photon trajectories in the vicinity
    of a black hole and its photon sphere. The blue trajectory represents
    weak deflection, for which the closest-approach radius satisfies
    $r_{0}\gg r_{\mathrm{m}}$. The orange trajectory illustrates strong
    deflection as $r_{0}\rightarrow r_{\mathrm{m}}^{+}$, where the photon
    winds around the black hole before escaping to infinity. The purple
    trajectory represents a captured photon that crosses the critical
    photon-sphere orbit and does not return to infinity.}
    \label{f2}
\end{figure}

\par
We now examine the singularity structure of the function $G(w)$, which
plays a central role in the subsequent analysis. In particular, the
location and nature of its singularities determine the analytic behaviour
of the transformed deflection integral and are therefore essential for
constructing a reliable approximation scheme.

We first define the photon-sphere radius $r_{\mathrm{m}}$. For a static,
spherically symmetric spacetime, the photon-sphere radius is determined
by the condition
\cite{Claudel2001,Bozza2002,Perlick2004}
\begin{equation}
\frac{C'(r_{\mathrm{m}})}{C(r_{\mathrm{m}})}
=
\frac{A'(r_{\mathrm{m}})}{A(r_{\mathrm{m}})},
\label{e40}
\end{equation}
where a prime denotes differentiation with respect to the radial
coordinate $r$.
Differentiating $h(r)=C(r)/A(r)$ with respect to $r$, we obtain
\begin{equation}
h'(r)
=
\frac{C'(r)A(r)-C(r)A'(r)}{A^{2}(r)}
=
\frac{C(r)}{A(r)}
\left[
\frac{C'(r)}{C(r)}
-
\frac{A'(r)}{A(r)}
\right].
\label{e41}
\end{equation}
Using the photon-sphere condition in Eq.~(\ref{e40}), the
quantity in square brackets vanishes at $r=r_{\mathrm{m}}$.
Consequently,
\begin{equation}
h'(r_{\mathrm{m}})=0.
\label{e42}
\end{equation}
We define the corresponding value of the transformed radial coordinate as
\begin{equation}
w_{\mathrm{m}}
\equiv h(r_{\mathrm{m}})
=
\frac{C(r_{\mathrm{m}})}{A(r_{\mathrm{m}})}.
\label{e43}
\end{equation}
Since $G(w)$, defined in Eq.~(\ref{e39}), contains $h'(r(w))$ in
the denominator, the vanishing of $h'(r)$ at the photon sphere implies
that $G(w)$ becomes singular as $w\rightarrow w_{\mathrm{m}}$.

An important distinction must be made between the radius of closest
approach $r_{0}$ and the photon-sphere radius $r_{\mathrm{m}}$. For the
scattering trajectories considered here, namely photons arriving from
infinity, reaching a radial turning point, and subsequently escaping
back to infinity, the physically relevant branch satisfies
\begin{equation}
r_{0}>r_{\mathrm{m}}.
\label{e44}
\end{equation}
Thus, $r_{\mathrm{m}}$ represents the limiting value of the closest
approach for the scattering branch and is approached only from above,
\begin{equation}
r_{0}\rightarrow r_{\mathrm{m}}^{+}.
\label{e45}
\end{equation}

The different possibilities are illustrated by the photon trajectories
shown in Fig.~\ref{f2}. When $r_{0}\gg r_{\mathrm{m}}$, represented by
the \textcolor{blue}{blue trajectory} in Fig.~\ref{f2}, the photon
remains far from the photon sphere and undergoes only a small change
in its propagation direction. This corresponds to the usual
weak-deflection regime.

As the closest-approach radius decreases towards the photon-sphere
radius, $r_{0}\rightarrow r_{\mathrm{m}}^{+}$, the behaviour changes
significantly. This case is represented by the
\textcolor{orange}{orange trajectory} in Fig.~\ref{f2}. The photon
approaches the unstable circular null orbit and spends an increasingly
long interval in its vicinity before eventually escaping to infinity.
Consequently, it can wind around the black hole one or more times,
and the accumulated azimuthal angle increases without bound. For a
nondegenerate unstable photon sphere, this behaviour gives rise to
the characteristic logarithmic divergence of the deflection angle
\cite{Darwin1959,Bozza2002,IyerPetters2007}.

At the critical value
\begin{equation}
r_{0}=r_{\mathrm{m}},
\qquad
w_{0}=w_{\mathrm{m}},
\label{e46}
\end{equation}
the radial turning point becomes degenerate, and the photon follows
the unstable circular null orbit at $r=r_{\mathrm{m}}$. This limiting
orbit separates the scattering and capture branches. On the capture
side of this critical orbit, represented by the
\textcolor{purple}{purple trajectory} in Fig.~\ref{f2}, the photon
has no outer turning point corresponding to the scattering integral
considered here and instead propagates towards the black-hole
interior. Therefore, $\mathcal{I}(r_{0})$, as defined for a photon
that returns to infinity, does not describe this captured trajectory.

The above behaviour also provides a physical interpretation of the
singularity of $G(w)$ at $w=w_{\mathrm{m}}$. As
$r_{0}\rightarrow r_{\mathrm{m}}^{+}$, the photon spends progressively
more time near the unstable circular null orbit, producing the
multiple-winding behaviour illustrated by the orange trajectory in
Fig.~\ref{f2}. The divergence of the deflection angle is therefore
directly associated with the photon-sphere structure rather than
being merely an artefact of the change of variables.

It is important, however, to distinguish this singularity from the
turning-point singularity appearing at the lower limit of the deflection
integral. For every scattering trajectory with
$r_{0}>r_{\mathrm{m}}$, the transformed integral contains the kernel
$(w-w_{0})^{-1/2}$. The singularity at $w=w_{0}$ is an integrable
square-root singularity associated with the ordinary radial turning
point and yields a finite value of $\mathcal{I}(w_{0})$ for fixed
$r_{0}>r_{\mathrm{m}}$. In contrast, the singularity of $G(w)$ at
$w=w_{\mathrm{m}}=h(r_{\mathrm{m}})$ originates from the geometric
condition $h'(r_{\mathrm{m}})=0$ and is therefore a property of the
spacetime geometry itself. As the lower limit approaches this
geometric singularity, $w_{0}\rightarrow w_{\mathrm{m}}$, the combined
behaviour of $G(w)$ and the Abel kernel produces the divergent
deflection characteristic of the photon-sphere limit.

To further characterize the photon-sphere singularity, we now examine
the local behaviour of $G(w)$ in the vicinity of
$w=w_{\mathrm{m}}$. The leading-order behaviour of $G(w)$ near this
critical point is established by the following lemma.
\begin{lemma}[Local behaviour of $G(w)$]
For a nondegenerate photon sphere at $r=r_{\mathrm{m}}$, satisfying
$h'(r_{\mathrm{m}})=0$ and $h''(r_{\mathrm{m}})>0$, the function $G(w)$
exhibits the following leading-order behaviour in the vicinity of
$w_{\mathrm{m}}=h(r_{\mathrm{m}})$:
\begin{equation}
G(w)\sim g_{0}(w-w_{\mathrm{m}})^{-1/2},
\qquad
w\rightarrow w_{\mathrm{m}}^{+},
\label{e47}
\end{equation}
where
\begin{equation}
g_{0}
=
\sqrt{\frac{B(r_{\mathrm{m}})}
{2C(r_{\mathrm{m}})h''(r_{\mathrm{m}})}}.
\label{e48}
\end{equation}
\end{lemma}

\begin{proof}
Since $r_{\mathrm{m}}$ is the photon-sphere radius, it satisfies
\begin{equation}
h'(r_{\mathrm{m}})=0.
\label{e49}
\end{equation}
We expand $h(r)$ about $r=r_{\mathrm{m}}$. Since the first derivative
vanishes at the photon sphere, the leading nonvanishing contribution
is quadratic:
\begin{equation}
h(r)-h(r_{\mathrm{m}})
\approx
\frac{1}{2}
h''(r_{\mathrm{m}})
(r-r_{\mathrm{m}})^{2}.
\label{e50}
\end{equation}
Using $w=h(r)$ and $w_{\mathrm{m}}=h(r_{\mathrm{m}})$, we obtain
\begin{equation}
w-w_{\mathrm{m}}
\approx
\frac{1}{2}
h''(r_{\mathrm{m}})
(r-r_{\mathrm{m}})^{2}.
\label{e51}
\end{equation}
Therefore, on the physically relevant outer branch
$r>r_{\mathrm{m}}$, we obtain
\begin{equation}
r-r_{\mathrm{m}}
\approx
\sqrt{
\frac{2(w-w_{\mathrm{m}})}
{h''(r_{\mathrm{m}})}
}.
\label{e52}
\end{equation}

We next expand $h'(r)$ in the neighbourhood of the photon sphere.
Since $h'(r_{\mathrm{m}})=0$, its leading-order behaviour is
\begin{equation}
h'(r)
\approx
h''(r_{\mathrm{m}})
(r-r_{\mathrm{m}}).
\label{e53}
\end{equation}
Substituting Eq.~(\ref{e52}) into Eq.~(\ref{e53}), we obtain
\begin{align}
h'(r(w))
&\approx
h''(r_{\mathrm{m}})
\sqrt{
\frac{2(w-w_{\mathrm{m}})}
{h''(r_{\mathrm{m}})}
}
\nonumber\\
&=
\sqrt{
2h''(r_{\mathrm{m}})
(w-w_{\mathrm{m}})
}.
\label{e54}
\end{align}
Substituting Eq.~(\ref{e54}) into the definition of $G(w)$
given in Eq.~(\ref{e39}), and evaluating the remaining metric
functions at $r=r_{\mathrm{m}}$, we obtain
\begin{align}
G(w)
&\approx
\frac{\sqrt{B(r_{\mathrm{m}})}}
{\sqrt{C(r_{\mathrm{m}})}
\sqrt{
2h''(r_{\mathrm{m}})
(w-w_{\mathrm{m}})
}}
\nonumber\\
&=
\sqrt{
\frac{B(r_{\mathrm{m}})}
{2C(r_{\mathrm{m}})h''(r_{\mathrm{m}})}
}
(w-w_{\mathrm{m}})^{-1/2}.
\label{e55}
\end{align}
Introducing
\begin{equation}
g_{0}
=
\sqrt{
\frac{B(r_{\mathrm{m}})}
{2C(r_{\mathrm{m}})h''(r_{\mathrm{m}})}
},
\label{e56}
\end{equation}
we therefore arrive at
\begin{equation}
G(w)
\sim
g_{0}(w-w_{\mathrm{m}})^{-1/2},
\qquad
w\rightarrow w_{\mathrm{m}}^{+}.
\label{e57}
\end{equation}

To express the coefficient $g_{0}$ explicitly in terms of the metric
functions, we differentiate
\begin{equation}
h(r)=\frac{C(r)}{A(r)}.
\label{e58}
\end{equation}
The first derivative is given by
\begin{equation}
h'(r)
=
\frac{
C'(r)A(r)-C(r)A'(r)
}
{A^{2}(r)}.
\label{e59}
\end{equation}
Differentiating once more, we obtain
\begin{equation}
\begin{aligned}
h''(r)
&=
\frac{
C''(r)A(r)-C(r)A''(r)
}
{A^{2}(r)}
\\
&\quad
-\frac{2A'(r)}{A^{3}(r)}
\left[
C'(r)A(r)-C(r)A'(r)
\right].
\end{aligned}
\label{e60}
\end{equation}
At $r=r_{\mathrm{m}}$, the photon-sphere condition implies
\begin{equation}
C'(r_{\mathrm{m}})A(r_{\mathrm{m}})
-
C(r_{\mathrm{m}})A'(r_{\mathrm{m}})
=0.
\label{e61}
\end{equation}
Consequently, the second term in Eq.~(\ref{e60}) vanishes,
yielding
\begin{equation}
h''(r_{\mathrm{m}})
=
\frac{
C''(r_{\mathrm{m}})A(r_{\mathrm{m}})
-
C(r_{\mathrm{m}})A''(r_{\mathrm{m}})
}
{A^{2}(r_{\mathrm{m}})}.
\label{e62}
\end{equation}
This completes the proof.
\end{proof}
\par
We now introduce a fixed  radius $r_1>r_{\mathrm{m}}$,
with $w_1=h(r_1)$. Motivated by Lemma 1, we separate the
leading singular contribution from $G(w)$ by defining
\begin{equation}
\widehat{G}(w)
=
G(w)-\frac{g_0}{\sqrt{w-w_{\mathrm{m}}}}.
\label{e63}
\end{equation}
From Lemma 1, it follows that $\widehat{G}(w)$ remains bounded
as $w\rightarrow w_{\mathrm{m}}^{+}$. Moreover, assuming sufficient
regularity of the metric functions, $\widehat{G}(w)$ admits a
continuous extension to $w=w_{\mathrm{m}}$. It is important to
emphasise that the decomposition
\begin{equation}
G(w)=\frac{g_0}{\sqrt{w-w_{\mathrm{m}}}}+\widehat{G}(w)
\label{e64}
\end{equation}
is an exact algebraic identity rather than an approximation.
Consequently, the leading singular contribution is explicitly
isolated without discarding any information about the original
function $G(w)$.

Having established the local singular structure of $G(w)$, we now
proceed to reformulate the deflection integral in a form that
separates the dominant photon-sphere contribution from the
remaining regular and asymptotic contributions. To this end,
we introduce the matching point $w_1=h(r_1)$, where
$r_1>r_{\mathrm m}$, and partition the integration domain at
$w=w_1$. Employing the exact decomposition of $G(w)$ derived
above, the deflection integral, for
$w_{\mathrm m}<w_0<w_1$, can be expressed as
\begin{equation}
\mathcal I(w_0)
=
\mathcal I_{\mathrm{sing}}(w_0)
+
\mathcal I_{\mathrm{reg}}(w_0)
+
\mathcal I_{\mathrm{far}}(w_0),
\label{e65}
\end{equation}
where
\begin{equation}
\mathcal I_{\mathrm{sing}}(w_0)
=
2g_0\sqrt{w_0}
\int_{w_0}^{w_1}
\frac{dw}
{\sqrt{(w-w_{\mathrm m})(w-w_0)}},
\label{e66}
\end{equation}
\begin{equation}
\mathcal I_{\mathrm{reg}}(w_0)
=
2\sqrt{w_0}
\int_{w_0}^{w_1}
\frac{\widehat G(w)}{\sqrt{w-w_0}}\,dw,
\label{e67}
\end{equation}
and
\begin{equation}
\mathcal I_{\mathrm{far}}(w_0)
=
2\sqrt{w_0}
\int_{w_1}^{\infty}
\frac{G(w)}{\sqrt{w-w_0}}\,dw.
\label{e68}
\end{equation}

This representation constitutes an exact partition of the
original integral, with no approximation introduced at this
stage. The first contribution explicitly isolates the
leading photon-sphere singularity and admits an analytical
evaluation. The second contains the bounded remainder
$\widehat G(w)$ and is therefore suitable for a systematic
Chebyshev--minimax treatment, whereas the third accounts
for the asymptotic region beyond the matching point.
Such a decomposition enables the distinct analytical
features of the deflection integral to be treated
independently, providing a natural basis for constructing
a controlled approximation scheme.

We first consider the singular contribution
$\mathcal I_{\mathrm{sing}}(w_0)$, which encodes the dominant
behavior of the deflection integral in the vicinity of the
photon sphere. By isolating the leading algebraic singularity
of $G(w)$, this contribution admits an exact analytical
evaluation without invoking any approximation. Explicitly,
\begin{equation}
\mathcal I_{\mathrm{sing}}(w_0)
=
2g_0\sqrt{w_0}
\int_{w_0}^{w_1}
\frac{dw}{\sqrt{(w-w_0)(w-w_{\mathrm m})}}.
\label{e69}
\end{equation}
The integral is elementary and can be expressed in logarithmic
form through the identity
\begin{align}
&\int_{w_0}^{w_1}
\frac{dw}{\sqrt{(w-w_0)(w-w_{\mathrm m})}}
\nonumber\\
&=
2\ln\left(
\sqrt{w_1-w_{\mathrm m}}
+
\sqrt{w_1-w_0}
\right)
-\ln(w_0-w_{\mathrm m}).
\label{e70}
\end{align}
Consequently, the singular contribution reduces exactly to
\begin{align}
\mathcal I_{\mathrm{sing}}(w_0)
=
2g_0\sqrt{w_0}\Bigg[
&2\ln\left(
\sqrt{w_1-w_{\mathrm m}}
+
\sqrt{w_1-w_0}
\right)
\nonumber\\
&-\ln(w_0-w_{\mathrm m})
\Bigg].
\label{e71}
\end{align}

The logarithmic structure explicitly characterizes the
asymptotic behavior of the deflection integral in the critical
limit $w_0\rightarrow w_{\mathrm m}^{+}$. In particular,
the divergent contribution is governed by
$-\ln(w_0-w_{\mathrm m})$, whereas the remaining terms
approach finite values. Thus, the photon-sphere divergence
is recovered analytically, with its leading coefficient
determined entirely by the local geometric quantities
entering $g_0$ \cite{Bozza2002,IyerPetters2007}.
\par
In the present approximation scheme, we employ a
Chebyshev--minimax approach to construct a systematic
approximation of the regular contribution to the
deflection integral. Before developing this construction,
we briefly review the relevant mathematical definitions
and fundamental properties of Chebyshev polynomials and
minimax approximation theory \cite{Trefethen2013,Boyd2001}.
\subsection{Chebyshev Polynomials: Definitions and Properties}
The Chebyshev polynomials of the first kind constitute a sequence
of orthogonal polynomials on the interval $[-1,1]$, defined through
the trigonometric representation \cite{Trefethen2013,Boyd2001}
\begin{equation}
T_n(x)=\cos\left(n\cos^{-1}x\right),
\qquad n=0,1,2,\ldots.
\label{e72}
\end{equation}
Introducing the transformation $x=\cos\theta$, with
$\theta\in[0,\pi]$, and employing the trigonometric identity
\begin{equation}
\cos[(n+1)\theta]+\cos[(n-1)\theta]
=2\cos\theta\cos(n\theta),
\label{e73}
\end{equation}
we obtain the three-term recurrence relation
\begin{equation}
T_{n+1}(x)=2xT_n(x)-T_{n-1}(x),
\label{e74}
\end{equation}
subject to the initial conditions
$T_0(x)=1$ and $T_1(x)=x$.
This recurrence establishes inductively that $T_n(x)$ is
a polynomial of degree $n$. In particular, the first few
members of the sequence are
\begin{align}
T_0(x)&=1, \qquad T_1(x)=x, \nonumber\\
T_2(x)&=2x^2-1, \nonumber\\
T_3(x)&=4x^3-3x, \nonumber\\
T_4(x)&=8x^4-8x^2+1, \nonumber\\
T_5(x)&=16x^5-20x^3+5x.
\label{e75}
\end{align}

\begin{figure}
    \centering
    \includegraphics[width=1.1\linewidth]{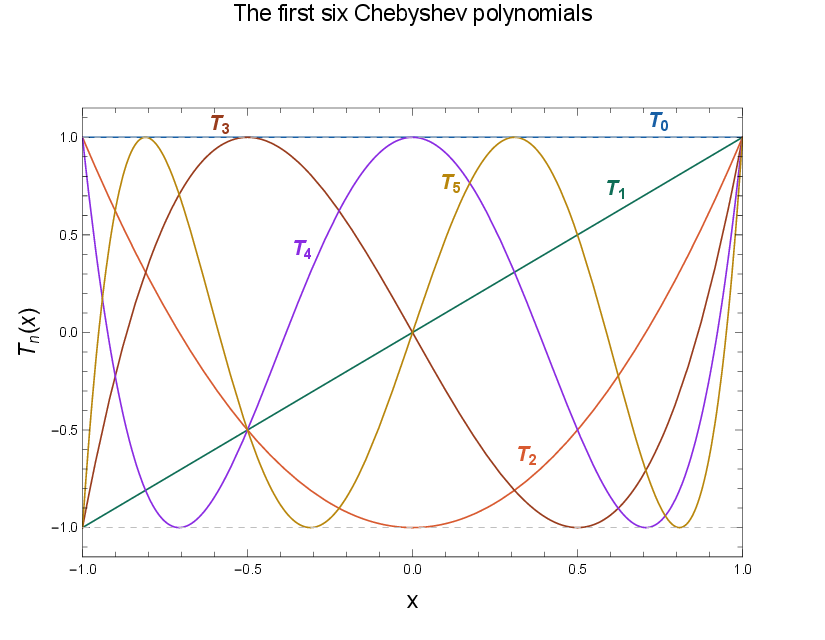}
    \caption{Plots of the first six Chebyshev polynomials
    $T_n(x)$, with $n=0,1,\ldots,5$, over the interval $[-1,1]$.}
    \label{f3}
\end{figure}

\par
\paragraph{Property 1: Uniform boundedness.}
From the trigonometric representation of the Chebyshev
polynomials and the bound $|\cos\theta|\leq 1$, it follows
immediately that
\begin{equation}
|T_n(x)|\leq 1,
\qquad x\in[-1,1],
\label{e76}
\end{equation}
for every nonnegative integer $n$. Thus, the Chebyshev
polynomials are uniformly bounded in absolute value,
independently of their degree. As illustrated in Fig.~\ref{f3},
each polynomial oscillates entirely within the interval $[-1,1]$.

\paragraph{Property 2: Equioscillation.}
A fundamental characteristic of Chebyshev polynomials is
their equioscillatory structure. Setting $x=\cos\theta$,
the extremal values of $T_n(x)=\cos(n\theta)$ occur when
$n\theta=k\pi$. Consequently, for $n\geq 1$, the
$n+1$ extremal points are given by
\begin{equation}
x_k=\cos\left(\frac{k\pi}{n}\right),
\qquad k=0,1,\ldots,n,
\label{e77}
\end{equation}
at which
\begin{equation}
T_n(x_k)=(-1)^k.
\label{e78}
\end{equation}
Hence, $T_n$ attains the values $+1$ and $-1$
alternately at $n+1$ distinct points of $[-1,1]$,
as illustrated in Fig.~\ref{f3}. This alternating extremal
structure underlies the Chebyshev alternation theorem,
which characterizes the polynomial of best uniform
approximation. Specifically, for a continuous function
$f$ on $[-1,1]$, a polynomial $p_N$ of degree at most
$N$ is the minimax approximant if and only if the error
$f-p_N$ attains its maximum absolute magnitude with
alternating signs at at least $N+2$ ordered points,
provided the error is nonzero. The polynomial
$T_{N+1}$ provides a canonical example of such an
alternating structure \cite{Trefethen2013,Rivlin1990}.

\par
\paragraph{Property 3: Roots and Chebyshev nodes.}
The roots of the Chebyshev polynomials follow directly
from their trigonometric representation. Setting
$T_n(x)=\cos(n\theta)=0$, with $x=\cos\theta$, yields
$n\theta=(2k+1)\pi/2$. Consequently, for $n\geq 1$,
the polynomial $T_n(x)$ possesses $n$ distinct roots
in the interval $(-1,1)$, given by
\begin{equation}
x_k=\cos\left(\frac{(2k+1)\pi}{2n}\right),
\qquad k=0,1,\ldots,n-1.
\label{e79}
\end{equation}
These roots define the Chebyshev--Gauss nodes, which
provide a nonuniform distribution of interpolation
points over $[-1,1]$. Unlike equally spaced nodes,
they become increasingly concentrated near the
endpoints $x=\pm 1$, while remaining relatively
sparse near the center. This distribution follows
from the projection $x=\cos\theta$ of uniformly
spaced angular points onto the real axis.

In the present approximation scheme, these nodes
will be employed to sample the regularized function
$\widehat G(w)$, after mapping the physical interval
$[w_{\mathrm m},w_1]$ onto $[-1,1]$. Their endpoint
clustering is particularly advantageous for resolving
variations near the boundaries and constructing
stable high-order polynomial approximations
\cite{Trefethen2013,Boyd2001}.

\par
\paragraph{Property 4: Orthogonality.}
The Chebyshev polynomials of the first kind form an
orthogonal sequence on $[-1,1]$ with respect to the
weight function $\omega(x)=(1-x^2)^{-1/2}$.
More precisely, they satisfy the orthogonality relation
\begin{equation}
\int_{-1}^{1}
\frac{T_n(x)T_m(x)}{\sqrt{1-x^2}}\,dx
=
\begin{cases}
0, & n\neq m,\\
\pi, & n=m=0,\\
\dfrac{\pi}{2}, & n=m\neq 0.
\end{cases}
\label{e80}
\end{equation}
This property follows directly from the substitution
$x=\cos\theta$, which transforms the integral into
\begin{equation}
\int_0^\pi
\cos(n\theta)\cos(m\theta)\,d\theta.
\label{e81}
\end{equation}
The orthogonality relation provides a natural basis
for expanding sufficiently regular functions in
Chebyshev series. In particular, a function $f(x)$
can be represented, under suitable convergence
conditions, as
\begin{equation}
f(x)=\sum_{n=0}^{\infty}a_nT_n(x),
\label{e82}
\end{equation}
where the coefficients are determined by
\begin{equation}
a_n=
\frac{2-\delta_{n0}}{\pi}
\int_{-1}^{1}
\frac{f(x)T_n(x)}{\sqrt{1-x^2}}\,dx,
\label{e83}
\end{equation}
with $\delta_{n0}$ denoting the Kronecker delta.
This orthogonal structure facilitates the systematic
construction of polynomial approximations to the
regularized function $\widehat G(w)$.

\par
\paragraph{Property 5: Extremal minimax property.}
A fundamental feature of Chebyshev polynomials is their
extremal property with respect to the uniform norm.
Among all monic polynomials of degree $n\geq 1$,
the rescaled Chebyshev polynomial
$2^{1-n}T_n(x)$ uniquely minimizes the maximum
absolute deviation on $[-1,1]$. More precisely,
\begin{equation}
\min_{\substack{p\in\mathcal P_n\\
p\text{ monic}}}
\|p\|_{\infty}
=
\left\|2^{1-n}T_n\right\|_{\infty}
=
2^{1-n},
\label{e84}
\end{equation}
where $\mathcal P_n$ denotes the space of
polynomials of degree at most $n$, and
\begin{equation}
\|p\|_{\infty}
=
\max_{x\in[-1,1]}|p(x)|.
\label{e85}
\end{equation}
Equivalently, every monic polynomial $p(x)$
of degree $n$ satisfies
\begin{equation}
\max_{x\in[-1,1]}|p(x)|
\geq
2^{1-n},
\label{e86}
\end{equation}
with equality if and only if
$p(x)=2^{1-n}T_n(x)$.

This extremal behaviour is a direct consequence of
the equioscillation property. The alternating
extrema of $T_n(x)$ distribute the maximum
deviation uniformly across $n+1$ distinct points,
preventing any further reduction of the uniform
norm while preserving the leading coefficient.
This principle forms the mathematical foundation
of the Chebyshev alternation theorem, which
characterizes the best uniform polynomial
approximation through the alternating extrema
of its error function. Consequently, Chebyshev
polynomials provide a natural basis for the
minimax treatment of the regularized function
$\widehat G(w)$ in the present formulation
\cite{Rivlin1990,Trefethen2013}.

\par
Having established the definition and fundamental properties of
Chebyshev polynomials, we now apply these results to approximate
the regularized function $\widehat{G}(w)$ arising in the lensing
problem. As a first step, we map its domain onto the canonical
Chebyshev interval $[-1,1]$.
\subsection{Chebyshev--Minimax Approximation of the Regularized Lensing Function}
\paragraph{Compactified variable.}
To implement the Chebyshev approximation, we first introduce
the inverse radial coordinate $u=1/r$. For a fixed matching
radius $r_1>r_{\mathrm m}$, the physical interval
$r\in[r_{\mathrm m},r_1]$ is mapped onto
$u\in[u_1,u_{\mathrm m}]$, where
\begin{equation}\label{e87}
u_1=\frac{1}{r_1},
\qquad
u_{\mathrm m}=\frac{1}{r_{\mathrm m}}.
\end{equation}
We then map this interval affinely onto the canonical
Chebyshev domain $[-1,1]$ through
\begin{equation}\label{e88}
x(u)=
\frac{2(u-u_1)}{u_{\mathrm m}-u_1}-1.
\end{equation}
Accordingly,
\begin{equation}\label{e89}
\begin{aligned}
r=r_1
&\;\longleftrightarrow\;
u=u_1
\;\longleftrightarrow\;
x=-1,\\
r=r_{\mathrm m}
&\;\longleftrightarrow\;
u=u_{\mathrm m}
\;\longleftrightarrow\;
x=1.
\end{aligned}
\end{equation}
Since $u=1/r$ reverses the radial ordering, the endpoint
$x=1$ corresponds to the photon sphere, whereas $x=-1$
corresponds to the outer matching radius. The subsequent
transformation $u\mapsto x$ is an order-preserving affine
rescaling, placing the physical domain precisely on the
standard interval required for the Chebyshev construction.

\paragraph{Regularized function in the Chebyshev variable.}
The regularized function was previously introduced as
\begin{equation}\label{e90}
\widehat{G}(w)
=
G(w)-\frac{g_0}{\sqrt{w-w_{\mathrm m}}}.
\end{equation}
To express this function on the standard Chebyshev interval, we define
\begin{equation}\label{e91}
\widetilde{G}(x)
\equiv
\widehat{G}\!\left(w(r(u(x)))\right).
\end{equation}
The corresponding sequence of transformations is therefore
\begin{equation}\label{e92}
x
\longrightarrow
u(x)
\longrightarrow
r(u)
\longrightarrow
w(r)
\longrightarrow
\widehat{G}(w).
\end{equation}
No new physical function is introduced through this transformation.
The quantities $\widetilde{G}(x)$ and $\widehat{G}(w)$ represent
the same regularized function expressed in different variables.
The former notation is introduced solely to formulate the
approximation on the canonical Chebyshev interval
$x\in[-1,1]$.

\paragraph{Chebyshev expansion.}
Since $\widetilde{G}(x)$ is regular on the interval under
consideration, it may be expanded in the Chebyshev basis as
\begin{equation}\label{e93}
\widetilde{G}(x)
=
\sum_{n=0}^{\infty}c_n T_n(x),
\end{equation}
where
\begin{equation}\label{e94}
T_n(x)
=
\cos\left(n\cos^{-1}x\right).
\end{equation}
For practical calculations, the infinite expansion is truncated
at a finite degree $N$, yielding
\begin{equation}\label{e95}
\widetilde{G}_N(x)
=
\sum_{n=0}^{N}c_n T_n(x).
\end{equation}
Thus, the approximation of $\widehat{G}(w)$ reduces to the
determination of the $N+1$ coefficients
$c_0,c_1,\ldots,c_N$.

\paragraph{Determination of the Chebyshev coefficients.}
The coefficients follow from the orthogonality relation
\begin{equation}\label{e96}
\int_{-1}^{1}
\frac{T_n(x)T_m(x)}
{\sqrt{1-x^2}}\,dx
=
\begin{cases}
0, & n\neq m,\\[1mm]
\pi, & n=m=0,\\[1mm]
\dfrac{\pi}{2}, & n=m\neq 0.
\end{cases}
\end{equation}
Multiplying the Chebyshev expansion by
$T_m(x)/\sqrt{1-x^2}$ and integrating over $[-1,1]$, we obtain
\begin{equation}\label{e97}
\int_{-1}^{1}
\frac{\widetilde{G}(x)T_m(x)}
{\sqrt{1-x^2}}\,dx
=
\sum_{n=0}^{\infty}
c_n
\int_{-1}^{1}
\frac{T_n(x)T_m(x)}
{\sqrt{1-x^2}}\,dx.
\end{equation}
By orthogonality, all terms with $n\neq m$ vanish, and hence
\begin{equation}\label{e98}
c_m
=
\frac{2-\delta_{m0}}{\pi}
\int_{-1}^{1}
\frac{\widetilde{G}(x)T_m(x)}
{\sqrt{1-x^2}}\,dx,
\end{equation}
where $\delta_{m0}$ denotes the Kronecker delta. Explicitly,
\begin{equation}\label{e99}
c_0
=
\frac{1}{\pi}
\int_{-1}^{1}
\frac{\widetilde{G}(x)}
{\sqrt{1-x^2}}\,dx,
\end{equation}
whereas, for $m\geq1$,
\begin{equation}\label{e100}
c_m
=
\frac{2}{\pi}
\int_{-1}^{1}
\frac{\widetilde{G}(x)T_m(x)}
{\sqrt{1-x^2}}\,dx.
\end{equation}

\paragraph{Fourier--cosine representation.}
Introducing
\begin{equation}\label{e101}
x=\cos\theta,
\qquad
0\leq\theta\leq\pi,
\end{equation}
together with
\begin{equation}\label{e102}
dx=-\sin\theta\,d\theta,
\qquad
\sqrt{1-x^2}=\sin\theta,
\end{equation}
and using
\begin{equation}\label{e103}
T_m(\cos\theta)=\cos(m\theta),
\end{equation}
the coefficient formula becomes
\begin{equation}\label{e104}
c_m
=
\frac{2-\delta_{m0}}{\pi}
\int_{0}^{\pi}
\widetilde{G}(\cos\theta)
\cos(m\theta)\,d\theta.
\end{equation}
Thus, the Chebyshev expansion can be interpreted as a
Fourier--cosine expansion under the transformation
$x=\cos\theta$.

\paragraph{Chebyshev--Gauss sampling.}
For numerical implementation, we evaluate
$\widetilde{G}(x)$ at the $N+1$ Chebyshev--Gauss nodes
\begin{equation}\label{e105}
x_k=\cos\theta_k,
\qquad
\theta_k=
\frac{(2k+1)\pi}{2(N+1)},
\qquad
k=0,1,\ldots,N.
\end{equation}
Equivalently,
\begin{equation}\label{e106}
x_k
=
\cos\left[
\frac{(2k+1)\pi}{2(N+1)}
\right].
\end{equation}
For each node $x_k$, the corresponding physical quantities
are obtained successively through
\begin{equation}\label{e107}
x_k
\longrightarrow
u_k
\longrightarrow
r_k
\longrightarrow
w_k
\longrightarrow
\widehat{G}(w_k),
\end{equation}
so that
\begin{equation}\label{e108}
\widetilde{G}(x_k)=\widehat{G}(w_k).
\end{equation}
Since $x_k\in(-1,1)$, neither endpoint is sampled directly.
In particular, the photon-sphere endpoint $x=1$ is avoided.

\paragraph{Discrete determination of the coefficients.}
The $(N+1)$-point Gauss--Chebyshev quadrature rule is
\begin{equation}\label{e109}
\int_{-1}^{1}
\frac{f(x)}{\sqrt{1-x^2}}\,dx
\approx
\frac{\pi}{N+1}
\sum_{k=0}^{N}f(x_k).
\end{equation}
Applying this relation to
$f(x)=\widetilde{G}(x)T_m(x)$ gives
\begin{equation}\label{e110}
c_m
\approx
\frac{2-\delta_{m0}}{N+1}
\sum_{k=0}^{N}
\widetilde{G}(x_k)T_m(x_k),
\qquad
m=0,1,\ldots,N.
\end{equation}
Since
\begin{equation}\label{e111}
T_m(x_k)=\cos(m\theta_k),
\end{equation}
the coefficients may equivalently be written as
\begin{equation}\label{e112}
c_m
\approx
\frac{2-\delta_{m0}}{N+1}
\sum_{k=0}^{N}
\widetilde{G}(x_k)\cos(m\theta_k).
\end{equation}
This expression is precisely a discrete cosine transform of
the sampled values $\widetilde{G}(x_k)$.

Equivalently, the coefficients $\{c_n\}_{n=0}^{N}$ may be
regarded as the solution of the system
\begin{equation}\label{e113}
\widetilde{G}(x_k)
=
\sum_{n=0}^{N}c_nT_n(x_k),
\qquad
k=0,1,\ldots,N.
\end{equation}
Thus, the degree-$N$ Chebyshev approximation is completely
determined once the regularized lensing function has been
evaluated at the Chebyshev--Gauss nodes.

\paragraph{Approximation Accuracy and Near-Minimax Property.}
\par
The accuracy and near-optimal character of the present Chebyshev
approximation are supported by two classical results of approximation
theory. \textbf{Jackson's theorem} guarantees that the best polynomial
approximation converges systematically as the polynomial degree $N$
increases, with the convergence rate controlled by the smoothness of
the approximated function. On the other hand, the \textbf{Chebyshev
equioscillation theorem} states that the minimax polynomial
$p_N^{\star}$ is characterized by an error that attains equal
maximum magnitude with alternating signs at at least $N+2$ points
of $[-1,1]$. Since the regularization removes the leading
photon-sphere singularity from $\widehat{G}$ and the Chebyshev basis
naturally possesses this extremal oscillatory structure, the resulting
Chebyshev approximation provides a stable and near-minimax
representation of the regularized lensing function.

\paragraph{Reconstruction in the original variable.}
Once $\widetilde{G}_N(x)$ has been constructed, we return to
the original variable and define
\begin{equation}\label{e114}
\widehat{G}_N(w)
=
\widetilde{G}_N(x(w)).
\end{equation}
Hence,
\begin{equation}\label{e115}
\widehat{G}(w)
\approx
\widehat{G}_N(w)
=
\sum_{n=0}^{N}
c_nT_n(x(w)).
\end{equation}
The regular contribution to the deflection integral therefore
takes the approximate form
\begin{equation}\label{e116}
\mathcal{I}_{\mathrm{reg}}(w_0)
\approx
2\sqrt{w_0}
\int_{w_0}^{w_1}
\frac{\widehat{G}_N(w)}
{\sqrt{w-w_0}}\,dw.
\end{equation}
Combining this result with the analytically isolated
photon-sphere contribution and the far-field contribution,
the degree-$N$ representation of the complete deflection
integral becomes
\begin{equation}\label{e117}
\begin{aligned}
\mathcal{I}_N(w_0)
={}&
2g_0\sqrt{w_0}
\int_{w_0}^{w_1}
\frac{dw}{\sqrt{(w-w_0)(w-w_{\mathrm m})}}\\
&+
2\sqrt{w_0}
\int_{w_0}^{w_1}
\frac{\widehat{G}_N(w)}{\sqrt{w-w_0}}\,dw\\
&+
2\sqrt{w_0}
\int_{w_1}^{\infty}
\frac{G(w)}{\sqrt{w-w_0}}\,dw .
\end{aligned}
\end{equation}
The three terms have distinct roles: the first contains the
photon-sphere singularity and is known analytically [Eq.~\eqref{e70}];
the second contains the regularized near-field contribution and
is treated by the Chebyshev approximation; and the third is a
smooth far-field integral that may be evaluated by standard
quadrature. In this manner, the singular and regular structures
of the lensing integral are separated before any numerical
approximation is introduced.
\section{Application to the Schwarzschild Black Hole}\label{Sec4}
In this section, we calculate the gravitational deflection angle of light in Schwarzschild spacetime using the fractional-calculus and Chebyshev approximation method developed in the preceding section. We derive the relevant expressions, evaluate the deflection angle, and examine the accuracy of the proposed method by comparing its results with direct numerical integration.
\par
\textbf{Step 1: Determination of the Photon-Sphere Radius}
\\
We consider the Schwarzschild spacetime, setting \(G=c=1\) and measuring all lengths in units of the black hole mass \(M\). Accordingly, we use the dimensionless radial coordinate \(r/M\), denoted simply by \(r\), so that \(r=3\) corresponds to \(r=3M\) in physical units. The metric functions are
\[
A(r)=1-\frac{2}{r},\qquad
B(r)=\frac{1}{A(r)},\qquad
C(r)=r^2.
\]

The photon-sphere radius \(r_{\mathrm m}\) is determined by the condition \(h'(r_{\mathrm m})=0\), where
\begin{equation}\label{e118}
  h(r)=\frac{C(r)}{A(r)}
=\frac{r^2}{1-\frac{2}{r}}
=\frac{r^3}{r-2}.  
\end{equation}

Differentiating \(h(r)\) with respect to \(r\), we obtain
\[
\begin{aligned}
h'(r)
&=\frac{3r^2(r-2)-r^3}{(r-2)^2}\\
&=\frac{2r^2(r-3)}{(r-2)^2}.
\end{aligned}
\]
Setting \(h'(r_{\mathrm m})=0\) gives
\[
2r^2(r-3)=0.
\]
Since the physical domain is \(r>2\), the only admissible solution is
\[
\boxed{r_{\mathrm m}=3.}
\]
Thus, the Schwarzschild photon sphere is located at \(r=3\) in dimensionless units, corresponding to \(r=3M\) when physical units are restored.
\par
 \textbf{Step 2: Determination of \(g_0\)}
 \\
The coefficient $(g_0)$, which characterises the leading-order behaviour of $(G(w))$ near the photon sphere, is given by
\[
g_0=\sqrt{\frac{B(r_m)}{2C(r_m)h''(r_m)}}.
\]
In dimensionless units, we set \(G=c=M=1\). The photon-sphere radius is therefore \(r_m=3\). From the Schwarzschild metric, we obtain
\[
h''(r_m)=18,\qquad B(r_m)=3,\qquad C(r_m)=r_m^2=9.
\]
Substituting the above values, we find
\[
g_0=\sqrt{\frac{3}{2\times9\times18}}
=\sqrt{\frac{3}{324}}
=\sqrt{\frac{1}{108}}.
\]
Hence,
\begin{equation}\label{e119}
\boxed{g_0=\frac{1}{6\sqrt{3}}\approx0.096225.}
\end{equation}
\par
\textbf{Step 3: Verification of Lemma 1}
We now verify Lemma 1, which states that near the photon sphere,
\begin{equation}\label{e120}
G(w)\approx \frac{g_0}{\sqrt{w-w_m}},
\qquad w\to w_m^+.
\end{equation}

For the Schwarzschild spacetime, using dimensionless units with \(M=1\), we have
\begin{equation}\label{e121}
h(r)=w=\frac{r^3}{r-2},
\qquad
w_m=h(r_m)=27,
\qquad r_m=3.
\end{equation}
The function \(G(w)\) can be expressed in terms of the radial coordinate as
\begin{equation}\label{e122}
G(r)=\frac{\sqrt{B(r)}}{\sqrt{C(r)}\,h'(r)}
=\frac{\sqrt{r}(r-2)^{3/2}}{2r^3(r-3)},
\qquad r>3.
\end{equation}

To examine the behaviour near the photon sphere, we expand \(h(r)\) about \(r=r_m=3\). Since \(h'(3)=0\) and \(h''(3)=18\), the leading-order expansion is
\begin{equation}\label{e123}
w-w_m=h(r)-h(3)
\approx \frac{1}{2}h''(3)(r-3)^2
=9(r-3)^2.
\end{equation}
Consequently,
\begin{equation}\label{e124}
\sqrt{w-w_m}\approx 3(r-3).
\end{equation}

Taking the limit of \(G(r)\sqrt{w-w_m}\) as \(r\to3^+\), we obtain
\begin{equation}\label{e125}
\begin{aligned}
\lim_{r\to3^+}G(r)\sqrt{w-w_m}
&=\lim_{r\to3^+}
\frac{\sqrt{r}(r-2)^{3/2}}{2r^3(r-3)}
\,3(r-3)\\
&=\frac{3\sqrt{3}}{2(27)}
=\frac{1}{6\sqrt{3}}.
\end{aligned}
\end{equation}

Using the result from Step 2, this limit is precisely \(g_0\). Therefore,
\begin{equation}\label{e126}
G(w)\sim\frac{1}{6\sqrt{3}\sqrt{w-27}},
\qquad w\to27^+.
\end{equation}

Thus, Lemma 1 is verified for the Schwarzschild spacetime in dimensionless units, with \(g_0=1/(6\sqrt{3})\) and \(w_m=27\).
\par
 \textbf{Step 4: Chebyshev Approximation of the Regularised Function}

In this step, we construct a Chebyshev polynomial approximation to the regularised function \(\widehat{G}(w)\). We choose the matching radius \(r_1=25\), while the photon-sphere radius is \(r_m=3\). To map the interval \(r\in[r_m,r_1]\) onto the standard Chebyshev interval \(x\in[-1,1]\), we introduce the transformation
\begin{equation}\label{e127}
u=\frac{1}{r}
=\frac{x+1}{2}\left(\frac{1}{r_m}-\frac{1}{r_1}\right)+\frac{1}{r_1},
\qquad r=\frac{1}{u}.
\end{equation}
Substituting \(r_m=3\) and \(r_1=25\), we obtain
\begin{equation}\label{e128}
u=\frac{11x+14}{75},
\qquad
r=\frac{75}{11x+14}.
\end{equation}

The regularised function is defined by subtracting the leading singular contribution from \(G(w)\):
\begin{equation}\label{e129}
\widehat{G}(w)=G(w)-\frac{g_0}{\sqrt{w-w_m}},
\end{equation}
where \(g_0=1/(6\sqrt{3})\) and \(w_m=27\). Using the transformation above, we express the regularised function in terms of the Chebyshev coordinate \(x\). For the Schwarzschild spacetime, this gives
\begin{equation}\label{e130}
\begin{aligned}
\widehat{G}(x)={}&
-\frac{1}{594}
\sqrt{-\frac{(-1+x)^2(53+22x)}
{(14+11x)^2(-47+22x)}}\\
&-\frac{(47-22x)^2(14+11x)}
{1856250(-1+x)}
\sqrt{-\frac{(14+11x)^2}{-141+66x}}.
\end{aligned}
\end{equation}

\begin{figure}
    \centering
    \includegraphics[width=1 \linewidth]{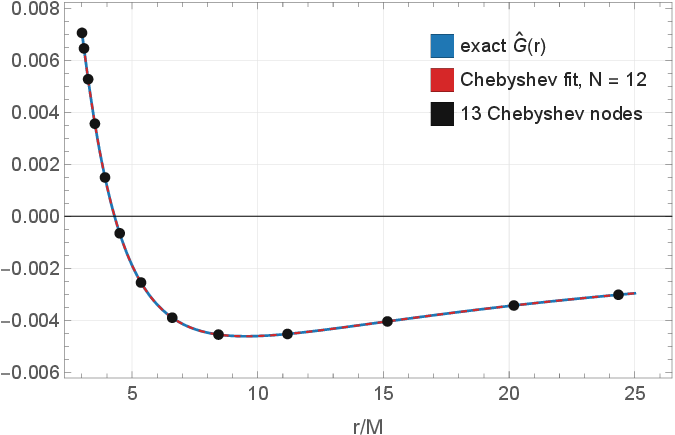}
    \caption{Comparison of the exact regularised function and its Chebyshev polynomial approximation of degree $N=12$ as functions of the radial coordinate $r$.}
    \label{f4}
\end{figure}

\begin{figure}
    \centering
    \includegraphics[width=1 \linewidth]{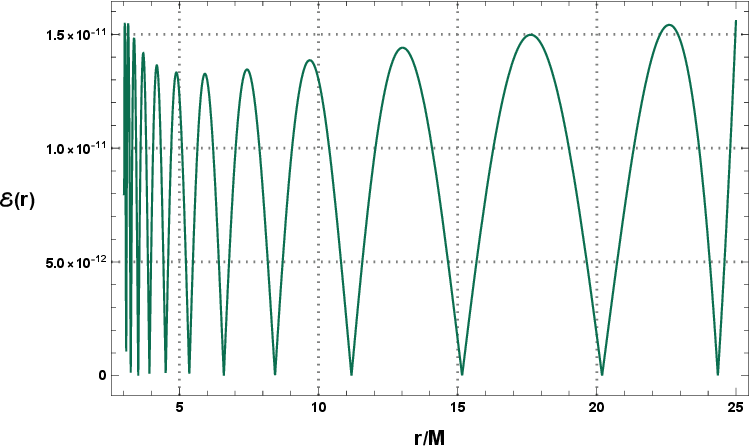}
    \caption{Absolute error between the exact regularised function and its Chebyshev polynomial approximation }
    \label{f5}
\end{figure}

We then approximate \(\widehat{G}(x)\) on \(x\in[-1,1]\) by a truncated Chebyshev series of degree \(N\):
\begin{equation}\label{e131}
\widehat{G}_N(x)=\sum_{n=0}^{N}c_nT_n(x),
\end{equation}
where \(T_n(x)\) denotes the Chebyshev polynomial of the first kind of degree \(n\), and \(c_n\) are the corresponding expansion coefficients. Returning to the variable \(w\), the approximation is written as
\begin{equation}\label{e132}
\widehat{G}_N(w)=
\sum_{n=0}^{N}c_nT_n\!\left(x(r(w))\right).
\end{equation}
This approximation will be used to evaluate the regular contribution to the deflection integral.
To determine the expansion coefficients \(c_n\), we employ the Chebyshev–Gauss nodes,
\begin{equation}\label{e133}
x_k=\cos\left(\frac{(2k+1)\pi}{2(N+1)}\right),
\qquad k=0,1,\ldots,N.
\end{equation}
The coefficients are calculated using the discrete Chebyshev expansion,
\begin{equation}\label{e134}
c_n=\frac{2}{N+1}\sum_{k=0}^{N}
\widehat{G}(x_k)\cos\left(\frac{n(2k+1)\pi}{2(N+1)}\right),
\qquad n=1,2,\ldots,N,
\end{equation}
while the zeroth coefficient is given by
\begin{equation}\label{e135}
c_0=\frac{1}{N+1}\sum_{k=0}^{N}\widehat{G}(x_k).
\end{equation}
Choosing the truncation order \(N=12\), we obtain the following coefficients:
\begin{table}[htbp]
\centering
\caption{Chebyshev expansion coefficients for $N=12$.}
\label{tab:chebyshev_coefficients}
\begin{tabular}{cc}
\hline
$n$ & $c_n$ \\
\hline
0  & $-2.20556\times10^{-4}$ \\
1  & $ 5.51832\times10^{-3}$ \\
2  & $ 2.31672\times10^{-3}$ \\
3  & $-4.65523\times10^{-4}$ \\
4  & $-1.47907\times10^{-5}$ \\
5  & $-6.07435\times10^{-6}$ \\
6  & $-1.06989\times10^{-7}$ \\
7  & $-1.97995\times10^{-7}$ \\
8  & $-5.5267\times10^{-10}$ \\
9  & $-7.73186\times10^{-9}$ \\
10 & $ 5.74164\times10^{-12}$ \\
11 & $-3.26707\times10^{-10}$ \\
12 & $-7.98926\times10^{-13}$ \\
\hline
\end{tabular}
\end{table}
Thus, the degree-12 Chebyshev approximation to the regularised function is
\begin{equation}\label{e136}
\widehat{G}_{12}(x)=\sum_{n=0}^{12}c_nT_n(x),
\qquad x\in[-1,1].
\end{equation}
This polynomial approximation is subsequently used to evaluate the regular part of the deflection integral. From Table \ref{tab:chebyshev_coefficients}, we observe a rapid decay in the magnitudes of the Chebyshev coefficients, with \(|c_n|\lesssim 10^{-7}\) for \(n\geq 6\), apart from small numerical variations. This behaviour suggests that the truncated expansion captures the regularised function reasonably well at relatively low polynomial order. We now transform the Chebyshev approximation \(\widehat{G}_{12}(x)\) back to the radial coordinate \(r\) using the inverse mapping \(x=x(r)\) and compare it with the exact regularised function \(\widehat{G}(r)\). The comparison is presented in Fig.\eqref{f4}, where the two curves overlap closely, indicating good agreement between the Chebyshev approximation and the exact function. The corresponding absolute error, defined as
\begin{equation}\label{e137}
\mathcal{E}(r)=\left|\widehat{G}_{12}(r)-\widehat{G}(r)\right|,
\end{equation}
is shown in Fig.\eqref{f5} and Table \eqref{T2}  to assess the accuracy of the approximation across the chosen radial interval.
\par
\begin{table}[htbp]
\centering
\caption{Absolute error between the exact and Chebyshev-approximated regularised functions.}
\label{T2}
\begin{tabular}{cc}
\hline
$r$ & $\mathcal{E}(r)$ \\
\hline
3.01  & $9.52257\times10^{-12}$ \\
4.01  & $7.09817\times10^{-12}$ \\
5.01  & $1.21425\times10^{-11}$ \\
6.01  & $1.29259\times10^{-11}$ \\
7.01  & $9.80322\times10^{-12}$ \\
8.01  & $7.95690\times10^{-12}$ \\
9.01  & $9.52291\times10^{-12}$ \\
10.01 & $1.29621\times10^{-11}$ \\
11.01 & $2.34403\times10^{-12}$ \\
12.01 & $9.69521\times10^{-12}$ \\
13.01 & $1.44200\times10^{-11}$ \\
14.01 & $1.05363\times10^{-11}$ \\
15.01 & $1.50148\times10^{-12}$ \\
16.01 & $7.92128\times10^{-12}$ \\
17.01 & $1.39174\times10^{-11}$ \\
18.01 & $1.45699\times10^{-11}$ \\
19.01 & $9.93283\times10^{-12}$ \\
20.01 & $1.66697\times10^{-12}$ \\
21.01 & $7.47012\times10^{-12}$ \\
22.01 & $1.41155\times10^{-11}$ \\
23.01 & $1.47043\times10^{-11}$ \\
24.01 & $5.76650\times10^{-12}$ \\
\hline
\end{tabular}
\end{table}
As shown in Fig. \eqref{f5} and Table \eqref{T2}, the absolute error is of the order of \(10^{-12}\) throughout the sampled radial interval. This indicates excellent numerical agreement between the exact regularised function and its degree-12 Chebyshev approximation, demonstrating the high accuracy of the approximation over the interval considered.
\par
\textbf{Step 5: Calculation of the Deflection Angle}

\begin{figure*}[t]
    \centering
    \includegraphics[width=0.8\textwidth]{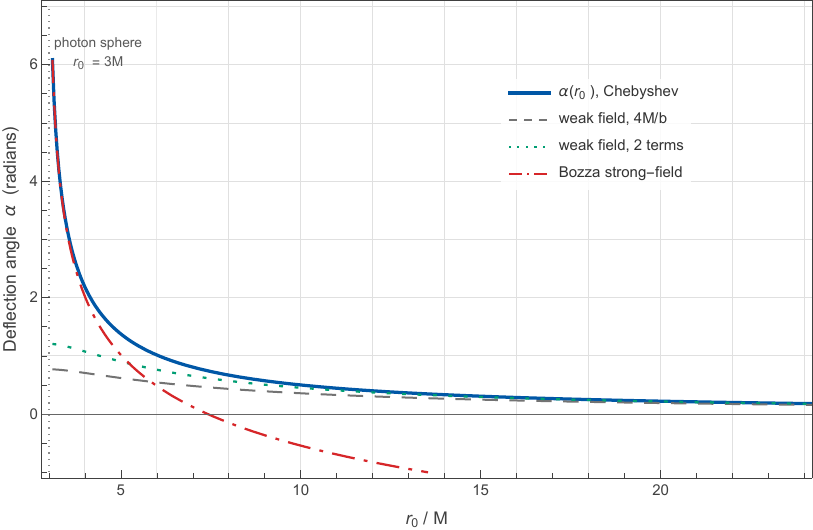}
    \caption{Caption}
    \label{f6}
\end{figure*}

In this step, we calculate the deflection angle for the Schwarzschild black hole using the results obtained in the preceding steps. Substituting the expressions for the singular, regular, and far-field contributions into Eq. \eqref{e117}, we evaluate the total deflection integral and subtract \(\pi\) to obtain the deflection angle. The resulting deflection angle is presented in Fig. \eqref{f6}. For comparison, we also plot the weak-deflection approximation and the result obtained using Bozza's method.
As shown in Fig. \eqref{f6}, our method provides a consistent description of the deflection angle over the radial interval considered. In contrast, the Bozza approximation may yield unphysical negative values sufficiently far from the photon sphere because it is based on a strong-deflection expansion designed for trajectories whose closest-approach radius approaches the photon-sphere radius, \(r_0\to r_m^+\). Such behaviour outside its intended regime of validity should not be interpreted as a physical prediction. The weak-deflection approximation is also included to compare the results in the regime of relatively large closest-approach radii.
A quantitative comparison of the deflection angles obtained using our method, the weak-deflection approximation, and Bozza's method is provided in Table \eqref{T3}.The table shows that the Chebyshev-based results agree closely with Bozza's strong-deflection approximation near the photon sphere. As \(r_0\) increases, the two results gradually diverge, reflecting the limited validity of the strong-deflection expansion away from the photon sphere. For \(r_0/M=8,15,\) and \(24\), Bozza's approximation yields negative deflection angles, whereas the Chebyshev results remain positive and approach zero as \(r_0\) increases. The two-term weak-deflection approximation also approaches the Chebyshev result at large \(r_0\), as expected in the weak-field regime.
\begin{table*}[htbp]
\centering
\caption{Comparison of the deflection angle obtained using the Chebyshev approximation, Bozza's strong-deflection approximation, and the two-term weak-deflection approximation for the Schwarzschild black hole. All quantities are expressed in dimensionless units with $M=1$.}
\label{T3}
\begin{tabular}{ccccc}
\hline
$r_0/M$ & $b/M$ & $\alpha$ (Chebyshev) & $\alpha$ (Bozza) & $\alpha$ (weak, 2 terms) \\
\hline
3.001 & 5.19615 & 15.2079 & 15.2079 & 1.20613 \\
3.01  & 5.19624 & 10.6108 & 10.6107 & 1.20611 \\
3.05  & 5.19823 & 7.42779 & 7.42644 & 1.20548 \\
3.1   & 5.20411 & 6.08632 & 6.08173 & 1.20362 \\
3.2   & 5.22558 & 4.78849 & 4.77358 & 1.19690 \\
3.5   & 5.34634 & 3.20612 & 3.14357 & 1.16034 \\
4.0   & 5.65685 & 2.18410 & 2.02269 & 1.07526 \\
5.0   & 6.45497 & 1.37674 & 1.01751 & 0.902421 \\
8.0   & 9.23760 & 0.669155 & $-0.148916$ & 0.571071 \\
15.0  & 16.1126 & 0.307264 & $-1.14258$ & 0.293632 \\
24.0  & 25.0672 & 0.181547 & $-1.74157$ & 0.178320 \\
\hline
\end{tabular}
\end{table*}

\section{Strong-field lensing observables from the Chebyshev deflection angle}
\label{sec5}

\subsection{Lens equation and relativistic images}

We consider a gravitational lensing configuration in which a source, a black hole acting as the lens, and an observer are nearly aligned. The observer--lens, lens--source, and observer--source distances are denoted by $D_{OL}$, $D_{LS}$, and $D_{OS}$, respectively, with $D_{OS}=D_{OL}+D_{LS}$. We assume that these distances are much larger than the characteristic length scale of the black hole, allowing us to work in the small-angle approximation. For a photon whose trajectory has a closest-approach radius $r_0$, the impact parameter in the Schwarzschild spacetime is
\begin{equation}\label{e138}
b=\frac{r_0}{\sqrt{1-2M/r_0}}.
\end{equation}
The corresponding deflection angle, $\alpha(b)$, is evaluated using the Chebyshev-based regularisation method developed in Sec.~\eqref{sec2}.

The image position $\theta$ and source position $\beta$ are related by the lens equation \cite{Virbhadra2000,Bozza2002}
\begin{equation}\label{e139}
\beta=\theta-\frac{D_{LS}}{D_{OS}}\Delta\alpha_n(\theta),
\qquad
\Delta\alpha_n=\alpha(b)-2\pi n,
\qquad
b\simeq D_{OL}\theta,
\end{equation}
where $n=1,2,\ldots$ denotes the number of complete revolutions made by the photon around the black hole. Since the deflection angle diverges logarithmically as the impact parameter approaches the critical value,
\begin{equation}\label{e140}
b_m=3\sqrt{3}\,M,
\end{equation}
the lens equation admits an infinite sequence of relativistic images. These images accumulate at the limiting angular position
\begin{equation}\label{e41b}
\theta_\infty=\frac{b_m}{D_{OL}}.
\end{equation}

\subsection{Image positions and magnifications}

For a source located exactly on the optical axis, $\beta=0$, the image positions are determined by
\begin{equation}\label{e142}
\alpha(b_n)=2\pi n,
\qquad
\theta_n^0=\frac{b_n}{D_{OL}}.
\end{equation}
For a small but nonzero source angle, we expand the deflection angle about the corresponding aligned-image solution. Defining
\begin{equation}\label{e143}
\alpha'_\theta
\equiv
D_{OL}\left.\frac{d\alpha}{db}\right|_{b=b_n},
\end{equation}
the first-order expansion of the lens equation gives
\begin{equation}
\theta_n\simeq\theta_n^0+
\frac{\beta-\theta_n^0}
{1+\dfrac{D_{LS}}{D_{OS}}\alpha'_\theta}.
\label{e144}
\end{equation}
The magnification of an image is determined by the Jacobian of the lens mapping. In the small-angle approximation, its signed value is
\begin{equation}
\mu_n=
\left[
\frac{\theta}{\beta}
\frac{d\theta}{d\beta}
\right]_{\theta=\theta_n}.
\label{e145}
\end{equation}
The absolute value $|\mu_n|$ gives the observable flux magnification. Using the linearised lens equation, we obtain
\begin{equation}
\mu_n\simeq
\frac{\theta_n}{\beta}
\left(
1+\frac{D_{LS}}{D_{OS}}\alpha'_\theta
\right)^{-1}.
\label{e146}
\end{equation}

The standard strong-field lensing observables are the limiting angular position $\theta_\infty$, the angular separation $s$ between the outermost relativistic image and the limiting position, and the magnitude difference $r_{\rm mag}$ between the first relativistic image and the remaining images \cite{Bozza2002,Bozza2010}:
\begin{align}
\theta_\infty
&=\frac{b_m}{D_{OL}}
=\frac{3\sqrt{3}\,M}{D_{OL}},
\label{e147}\\
s&=\theta_1-\theta_\infty,
\label{e148}\\
r_{\rm mag}
&=2.5\log_{10}
\left(
\frac{|\mu_1|}{\displaystyle\sum_{n\geq2}|\mu_n|}
\right).
\label{e149}
\end{align}

\subsection{Evaluation using the Chebyshev deflection angle}

In the conventional strong-deflection approach, the deflection angle is approximated by its leading logarithmic form,\cite{Bozza2002,Bozza2010}
\begin{equation}\label{e150}
\alpha(b)\simeq
-a\ln\left(\frac{b}{b_m}-1\right)+\bar b,
\end{equation}
where, for the Schwarzschild spacetime,
\begin{equation}\label{e151}
a=1,
\qquad
\bar b=\ln\left[216(7-4\sqrt{3})\right]-\pi
\simeq-0.4002.
\end{equation}
This approximation yields the standard expressions
\begin{equation}
s_{\rm B}=\theta_\infty
\exp\left(\frac{\bar b-2\pi}{a}\right),
\qquad
r_{\rm mag}^{\rm B}=\frac{5\pi}{a\ln 10}.
\label{e152}
\end{equation}

In the present approach, we instead evaluate the deflection angle directly from the Chebyshev approximation, without imposing a logarithmic functional form over the entire domain. The relativistic-image positions and magnifications are obtained through the following procedure:
\begin{enumerate}
\item For each winding number $n$, solve
\begin{equation}\label{e153}
\alpha(r_0)=2\pi n
\end{equation}
numerically for $r_0$. To improve numerical stability near the photon sphere, we introduce the auxiliary variable $x=\sqrt{r_0-3M}$ and use an initial estimate motivated by the strong-deflection expansion.

\item Calculate the corresponding impact parameter,
\begin{equation}\label{e154}
b_n=\frac{r_0}{\sqrt{1-2M/r_0}}.
\end{equation}
Because $b_n$ approaches $b_m$ rapidly as $n$ increases, direct subtraction may cause a loss of numerical precision. We therefore evaluate their difference using the identity
\begin{equation}\label{e155}
b_n-b_m=
\frac{(r_0-3M)^2(r_0+6M)}
{(r_0-2M)(b_n+b_m)}.
\end{equation}

\item Evaluate the derivative $d\alpha/db$ numerically from the Chebyshev deflection function. The chain rule gives
\begin{equation}\label{e156}
\frac{d\alpha}{db}
=
\frac{d\alpha/dx}{db/dx},
\qquad
\frac{db}{dx}=\frac{x\,h'(r_0)}{b},
\qquad
h(r)=\frac{r^3}{r-2M}.
\end{equation}

\item Substitute the resulting image positions and derivatives into Eqs.~\eqref{e147}--\eqref{e149} to calculate the angular positions, magnifications, and strong-field observables.
\end{enumerate}

This procedure allows the lensing observables to be evaluated directly from the numerically reconstructed deflection angle. It also provides a means of assessing the accuracy of the standard logarithmic approximation in the regime where that approximation is expected to be valid.

\subsection{Results}

We illustrate the method using the supermassive black hole at the Galactic centre, Sgr~A$^*$\cite{EHT2019M87,EHT2022SgrA}, with
\begin{equation}\label{e157}
\begin{aligned}
M &= 4.3\times10^6\,M_\odot, &\qquad D_{OL} &= 8.3~\mathrm{kpc},\\
D_{OS} &= 2D_{OL},         &\qquad \beta  &= 0.1\,\theta_\infty.
\end{aligned}
\end{equation}
For this configuration, the limiting angular position is
$\theta_\infty=26.571~\mu\mathrm{as}$. The first four relativistic images calculated using the degree-12 Chebyshev approximation with matching radius $r_1=25M$ are listed in Table~\ref{T4}.

\begin{table}[htbp]
\centering
\caption{First four relativistic images of Sgr~A$^*$ calculated using the Chebyshev deflection angle with $N=12$ and $r_1=25M$.}
\label{T4}
\small
\setlength{\tabcolsep}{3pt}
\resizebox{\columnwidth}{!}{%
\begin{tabular}{cccccc}
\hline\hline
$n$ & $(r_0-3M)/M$ & $(b_n-b_m)/M$ &
$\theta_n-\theta_\infty$ ($\mu$as) &
$|\alpha'_\theta|$ & $|\mu_n|$ \\
\hline
1 & $9.023\times10^{-2}$ & $6.528\times10^{-3}$ & $3.338\times10^{-2}$ & $6.16\times10^{12}$ & $3.252\times10^{-12}$ \\
2 & $3.751\times10^{-3}$ & $1.214\times10^{-5}$ & $6.21\times10^{-5}$ & $3.32\times10^{15}$ & $6.021\times10^{-15}$ \\
3 & $1.618\times10^{-4}$ & $2.268\times10^{-8}$ & $1.16\times10^{-7}$ & $1.78\times10^{18}$ & $1.124\times10^{-17}$ \\
4 & $6.99\times10^{-6}$ & $4.2\times10^{-11}$ & $2.2\times10^{-10}$ & $9.5\times10^{20}$ & $2.10\times10^{-20}$ \\
\hline\hline
\end{tabular}%
}
\end{table}

The first four relativistic images approach the limiting angular position as the winding number increases, while their magnifications decrease rapidly. This trend is consistent with the accumulation of relativistic images near the critical impact parameter\cite{Darwin1959,VirbhadraEllis2000,Bozza2010}.

Table~\ref{T4} compares the strong-field observables obtained from the Chebyshev deflection angle with those predicted by the standard logarithmic approximation.

\begin{table}[htbp]
\centering
\caption{Comparison of the strong-field lensing observables obtained using the Chebyshev method and the standard strong-deflection approximation.}
\label{T5}
\begin{tabular}{lccc}
\hline\hline
Observable & Chebyshev & Bozza & Difference \\
\hline
$\theta_\infty$ (Sgr~A$^*$) & $26.571~\mu$as & $26.571~\mu$as & $0$ \\
$s$ (Sgr~A$^*$) & $0.033381~\mu$as & $0.033253~\mu$as & $+0.38\%$ \\
$\theta_\infty$ (M87$^*$) & $19.844~\mu$as & $19.844~\mu$as & $0$ \\
$s$ (M87$^*$) & $0.024930~\mu$as & $0.024834~\mu$as & $+0.38\%$ \\
$r_{\rm mag}$ & $6.8292$ mag & $6.8219$ mag & $+0.0073$ mag \\
\hline\hline
\end{tabular}
\end{table}

The comparison indicates close agreement between the Chebyshev-based results and the standard strong-deflection predictions for the Schwarzschild spacetime. The relative difference in the angular separation is approximately $0.38\%$, while the difference in the magnitude observable is approximately $0.0073$ mag. These results demonstrate that the Chebyshev procedure reproduces the standard strong-field observables with high accuracy for this benchmark spacetime, while allowing the deflection angle to be evaluated without assuming a logarithmic form throughout the calculation.

The angular separation of the first relativistic image from the limiting position is approximately $0.03~\mu$as, and the image magnification is extremely small. Resolving these individual relativistic images therefore remains challenging with current very-long-baseline interferometric observations\cite{Bozza2010,EHT2019M87,EHT2022SgrA}.. By contrast, the limiting angular position $\theta_\infty$ provides the characteristic angular scale of the black-hole shadow and is more directly connected to current horizon-scale imaging observations\cite{EHT2019M87,EHT2022SgrA}.

\section{Conclusion}\label{Sec6}

In this work, we have developed a fractional-calculus-based framework for evaluating the gravitational deflection of light in static, spherically symmetric spacetimes. By introducing a suitable transformation of the radial coordinate, the deflection integral was recast into an Abel-type integral equation, allowing the underlying kernel to be reconstructed through fractional integration. A central feature of the proposed approach is the analytical isolation of the leading singular contribution near the photon sphere. Subtracting this contribution yields a regularised function that can be efficiently approximated using Chebyshev polynomials.

(i) We have employed Chebyshev--Gauss nodes to construct a truncated polynomial approximation of the regularised kernel. This procedure provides an accurate numerical representation over a finite radial interval while retaining the analytically known singular behaviour. The Schwarzschild spacetime was considered as a benchmark to examine the accuracy and reliability of the method. The numerical comparison between the exact regularised function and its Chebyshev approximation demonstrates excellent agreement [Fig.\eqref{f4}\eqref{f5}], with the absolute error reaching approximately $10^{-12}$ for the degree-12 approximation considered \eqref{T2}.

(ii) Using the reconstructed kernel, we evaluated the deflection angle by combining the singular, regular, and far-field contributions. The resulting deflection angle was compared with the standard strong-deflection approximation of Bozza and the two-term weak-deflection approximation [Fig.\eqref{f6}]. The Chebyshev-based results reproduce the expected behaviour near the photon sphere and remain well behaved at larger closest-approach radii. In particular, the comparison illustrates the limitations of extrapolating the logarithmic strong-deflection approximation beyond its intended regime of validity.

(iii) Furthermore, we applied the resulting deflection angle to investigate strong-field gravitational lensing observables, including the limiting angular position, the angular separation of relativistic images, and their relative magnifications. As an illustrative application, we considered the supermassive black holes Sgr~A$^*$ . For the Schwarzschild benchmark, the observables obtained using the Chebyshev procedure show close agreement with the standard strong-deflection predictions, with a relative difference of approximately $0.38\%$ in the angular separation and a difference of approximately $0.0073$ mag in the magnitude observable.

(iv) The present framework therefore provides an accurate and flexible alternative for evaluating gravitational deflection angles without imposing a logarithmic approximation over the entire domain of integration. Although the Schwarzschild spacetime serves as the benchmark in this work, the formulation is applicable to a broader class of static, spherically symmetric geometries, provided that the relevant photon-sphere structure and asymptotic behaviour are appropriately incorporated. Its extension to black holes in modified theories of gravity and other non-Schwarzschild geometries offers a promising direction for future investigations.

\section*{Acknowledgements}
AG is thankful to IIEST, Shibpur, India, for providing an Institute
Fellowship (SRF).

\bibliographystyle{naturemag}
\bibliography{bibliography}
\end{document}